\documentstyle[amsfonts,amstex,aps,epsfig]{revtex}
\begin{document}

\begin{center}
{\bf {The charm quark EDM and singlet $P$-wave charmonium production in
supersymmetry}}

\bigskip

Z. Z. Aydin and U. Erkarslan

\bigskip

Ankara University, Faculty of Engineering,\\[0pt]
Department of Engineering Physics, 06100 Tandogan, Ankara--TURKEY \bigskip
\end{center}
\begin{abstract}
We analyze the singlet $P$--wave charmonium production at $e^+
e^-$ colliders within the framework of unconstrained
supersymmetry. We show that the CP--violating transitions,
dominated by the gluino exchange, are typically four orders of
magnitude larger than the CP--conserving ones, and former is
generated by the electric dipole moment of the charm quark. Our
results can be directly tested via the charmonium searches at the
CLEO--c experiment.
\end{abstract}

\vspace{0.5cm}

{\bf I. INTRODUCTION}

\bigskip

A permanent electric dipole moment (EDM) of an elementary particle is a
clear signature for the CP violation. In the Standard Model (SM) of
electroweak interactions CP-violation originates from the phase in the CKM
matrix; but in this model the single fermion's EDM vanishes up to two-loop
order, and also three-loop contributions partially cancel. Therefore, the SM
contribution is neither relevant phenomenologically nor numerically
important. However, the Supersymmetric (SUSY) extension of SM contains many
new sources of CP violation generated by soft SUSY breaking terms. Actually
such a SUSY Model has many phases, but not all of them are physical. After
using some symmetries such as R-symmetry and the Peccie-Quinn symmetry, only
two of them remain physical: the trilinear coupling A and the $\mu $
parameter.

The upper bounds for the neutron and electron EDMs imply that CP violation
phases are duly small if the exchanged SUSY particles have masses close to
their current experimental lower limits. There are several suggestions to
find the electron and neutron EDMs below the experimental upper limits when
CP phases $\sim $O(1). Two possibilities have been commonly discussed in the
literature which include (i ) finding appropriate parameter domain where
different contributions cancel or (ii)making the first two generations of
scalar fermions heavy enough but keeping the soft masses of \ the third
generation below the TeV scale.

As the analysis of Ref. \cite{demvol} shows clearly, the electric dipole
moments (EDM) of the heavy quarks play an important role in the direct
production of singlet P--wave mesons at $e^{+}e^{-}$ colliders. Indeed, it
is possible to observe ${}^{1}\mbox{P}_{1}$ bottomonium in $B$ factories
like BaBaR, KEK or BEPC with sufficiently high statistics provided that the $%
b$ quark possesses a large enough EDM. Since the single fermion EDMs are
exceedingly small in the standard electroweak theory \cite{shahida} the SUSY
contribution remains as the only viable option. Indeed, in SUSY the EDMs
exist already at the one-loop level and typically exceed the existing $%
1.5\sigma $ upper bounds.

However, not only the bottom quark but also the charm quark EDM offers an
important arena in searching for the singlet P--wave charmonia. Indeed, in
near future with increasing data at CLEO-c experiment \cite{cleo-c}, it
might be possible to produce the ${}^{1}\mbox{P}_{1}$ charmonium state, $%
h_{c}$, directly. Therefore, a direct estimate of the charm quark EDM ${\cal
{D}}_{c}$ in SUSY will help in predicting how large the effect could be at
CLEO-c.

In our analysis we will concentrate on unconstrained low-energy SUSY model
as in Ref. \cite{demvol} in that portion of the parameter space where large
SUSY contributions to electron and neutron EDMs are naturally suppressed.
Such a region of the SUSY parameter space is characterized by a small phase
for the $\mu $ parameter: $-\pi /(5\tan \beta )\leq \varphi _{\mu }\leq \pi
/(5\tan \beta )$ \cite{cancel}.

Assuming universality of the gaugino masses at some ultra high scale, it is
known that the only physical soft phases are those of the $A$ parameters, $%
\varphi _{A_{i}}$ ($i=u,c,\cdots b$) and the $\mu $ parameter. Therefore, in
what follows we will scan the SUSY parameter space by varying ($i$) all soft
masses from $m_{t}$ to $1\ {\rm TeV}$, ($ii$) the phases of the $A$
parameters from $0$ to $\pi $, and finally, ($iii$) the phase of the $\mu $
parameter in bounds noted above suggested by the cancellation mechanism.

In the second section, we calculate the chargino and gluino contributions to
the c-quark EDM in effective SUSY at one-loop level and show that the gluino
contribution dominates.

Section 3 is devoted to the analysis of our numerical calculations.

In section 4, we discuss the possible signatures of the c-quark EDM in the $%
e^{+}e^{-}$ annihilation. After examining the formation of $^{1}P_{1}$
charmonium state in $e^{+}e^{-}$ channel, we show that the CP-violating
part, generated by the c-quark EDM dominates over the CP-conserving ones. In
the case of observing the CP-odd resonance \ in the data accumulating at
CLEO-c experiment, this would be a direct \ evidence for the charm quark's
EDM and also for the CP- violation in SUSY.

\bigskip

\bigskip

\bigskip

{\bf 2.THE CHARM QUARK EDM in SUSY}

\bigskip

The EDM of a spin 1/2 particle is defined as follows

\begin{equation}
{\frak L}=-\frac{i}{2}{\cal {D}}_{f}\overline{\psi }\sigma _{\mu \nu }\gamma
_{5}\psi F^{\mu \nu }
\end{equation}
which obviously appears as a quantum loop-effects. In the SM, ${\cal {D}}%
_{f} $ vanishes up to two-loop order \cite{shabalin}, whereas in SUSY it
exists already at one-loop level.

As shown in Fig. 1 the $c$ EDM is generated by the gluino, chargino and
neutralino diagrams at one loop level. The general expressions for the quark
EDMs in SUSY can be found in \cite{tarek}. At large $\tan \beta $ the
individual contributions to the $c$ quark EDM take the following compact
forms. The gluino contribution is given by
\begin{eqnarray}
\left( \frac{{\cal {D}}_{c}}{e}\right) ^{\tilde{g}} &=&\left( \frac{\alpha
_{s}(M_{SUSY})}{\alpha _{s}(m_{t})}\right) ^{16/21}\ \left( \frac{\alpha
_{s}(m_{t})}{\alpha _{s}(m_{b})}\right) ^{16/23}\ \left( \frac{\alpha
_{s}(m_{b})}{\alpha _{s}(m_{c})}\right) ^{\frac{16}{25}}\frac{2\alpha _{s}}{%
3\pi }\,Q_{c}\,\frac{m_{\widetilde{g}}m_{c}\mbox{Im}[A_{c}^{\ast }]}{M_{%
\tilde{c}_{1}}^{2}-M_{\tilde{c}_{2}}^{2}}  \nonumber  \label{gluana} \\
&\times &\left[ \frac{\mbox{B}(m_{\widetilde{g}}^{2}/M_{\tilde{c}_{1}}^{2})}{%
M_{\tilde{c}_{1}}^{2}}-\frac{\mbox{B}(m_{\widetilde{g}}^{2}/M_{\tilde{c}%
_{2}}^{2})}{M_{\tilde{c}_{2}}^{2}}\right]
\end{eqnarray}
where the loop function $\mbox{B}(x)$ can be found in \cite{tarek}. Here $M_{%
\tilde{c}_{1,2}}^{2}$ are the masses of the charm squark. It is clear that
at large $\tan \beta $ the (dominant) gluino--scharm contribution to $c$%
--quark EDM is almost independent of the phase of the $\mu $ parameter.

The chargino contribution, which proceeds via the exchange of charged
gauginos and Higgsinos together with the scalar $s$--quark does have a
direct dependence on the phase of the $\mu $ parameter via chargino mass
matrix \cite{demvol}. The exact expression reads as
\begin{eqnarray}
\left( \frac{{\cal {D}}_{c}}{e}\right) ^{\chi ^{\pm }} &=&-
\left( \frac{%
\alpha _{s}(M_{SUSY})}{\alpha _{s}(m_{t})}\right) ^{16/21}\left( \frac{%
\alpha _{s}(m_{t})}{\alpha _{s}(m_{b})}\right) ^{16/23}\ \left( \frac{\alpha
_{s}(m_{b})}{\alpha _{s}(m_{c})}\right) ^{\frac{16}{25}}  \nonumber \\
&&\times \frac{\alpha _{2}}{4\pi }\,Q_{c}\sum_{k=1}^{2}\sum_{j=1}^{2}\Im %
\left[ \Gamma _{\chi ^{\pm }}^{kj}\right] \,\frac{1}{M_{\chi _{j}^{\pm }}}%
\,F_{\pm }\left( \frac{M_{\chi _{j}^{\pm }}^{2}}{M_{\tilde{s}_{k}}^{2}}%
\right)   \nonumber \\
&=&-\left( \frac{\alpha _{s}(M_{SUSY})}{\alpha _{s}(m_{t})}\right)
^{16/21}\left( \frac{\alpha _{s}(m_{t})}{\alpha _{s}(m_{b})}\right)
^{16/23}\ \left( \frac{\alpha _{s}(m_{b})}{\alpha _{s}(m_{c})}\right) ^{%
\frac{16}{25}}  \nonumber \\
&&\times \frac{\alpha _{2}}{4\pi }\,\frac{m_{c}}{M_{W}^{2}}\ \Im \left[
C_{7}^{\chi ^{\pm }}(M_{W})\right]
\end{eqnarray}
where the first line results from the direct computation, and depends on the
vertex factors $\Gamma _{\chi ^{\pm }}^{kj}$ and the loop function $F_{\pm }$
both defined in Ref. \cite{demvol}. The second line, however, is the dipole
coefficient arising in $c\rightarrow u\gamma $ decay as was pointed out in
Ref. \cite{demvol} for $b\rightarrow s\gamma $. (This relation to rare
radiative decays is particularly important as $C_{7}^{\chi ^{\pm }}(M_{W})$
is particularly sensitive to large $\tan \beta $ effects \cite{demolive}
which introduces large enhancements compared to the SM.) A precise
determination of the branching ratio and the CP--asymmetry in $c\rightarrow
u\gamma $(or the hadronic mode $D\rightarrow \pi \gamma $ ) \cite{boz} decay
will help in fixing the chargino contribution to the $c$--quark EDM.
Presently, the experimental data is not precise enough to bound this
contribution so that we will analyze the result of the direct calculation in
(3) in the numerical studies below.

The neutralino contribution is at least 20 times smaller numerically than
the gluino contribution. So we do not include it in our numerical analysis
below.

\bigskip

{\bf 3.NUMERICAL ANALYSIS}

\bigskip We now discuss numerically the gluino and chargino contributions to
the EDM of the c-quark. As mentioned before, in our analysis we take the
SUSY parameters in the following ranges:

\begin{itemize}
\item  Soft masses: $m_{t}^{2}\leq M_{\tilde{s}_{k}}^{2},M_{\tilde{c}%
_{k}}^{2},|A_{c,s}|^{2}\leq \left( 1\ {\rm TeV}\right) ^{2}$

\item  Gaugino Masses: $m_{t}\leq M_{2}\leq 1\ {\rm TeV}$ $\mbox{with}$ $m_{%
\widetilde{g}}=\frac{\alpha _{s}}{\alpha _{2}}M_{2}$

\item  Phase of the $\mu $ parameter: $-\frac{\pi }{5\tan \beta }\leq
\varphi _{\mu }\leq \frac{\pi }{5\tan \beta }$

\item  Phase of the $A$ parameters: $0\leq \varphi _{A_{c,s}}\leq \pi $

\item  $\tan \beta $: $10\leq \tan \beta \leq 50$
\end{itemize}

where the modulus of $\mu $ is determined via the relation
\begin{eqnarray}
\left| A_{c}-\mu \cot \beta \right| =\frac{1}{4m_{c}^{2}}\left[ \left( M_{%
\tilde{c}_{1}}^{2}-M_{\tilde{c}_{2}}^{2}\right) ^{2}-rM_{W}^{4}(\cos 2\beta
)^{2}\right]
\end{eqnarray}
with $r=(g_{2}^{2}-(5/3)g_{1}^{2})/(2g_{2}^{2})$.
%-----------------------------------------------------------------------

In Fig.2 we plot the gluino contribution, ${\cal {D}}_{c}^{\widetilde{g}},$
in units of 10$^{-26}$e.cm, as a function of the A$_{c}$ phase (Fig.2(a)), $%
\mu $ parameter phase (Fig.2(c)) and tan$\beta $ (Fig.2(b)). We have to
emphasize that here we use the exact formula instead of Eq.(2). In all these
plots, we scan the other parameters in their allowed ranges. It is clear
from Fig. 2(a) that for most of the $\varphi _{A}$ space (i.e. from 0 to $%
\pi $) the value of the EDM is around 10$^{-21}$ e.cm, showing a
slight maximum at $\pi /2$. At $\varphi _{A}$=0, the 10$^{2}$
times smaller value of ${\cal {D}}_{c}^{\widetilde{g}}$ comes from
the $\varphi _{\mu }$ phase in the large tan$\beta $ limit. As we
have already mentioned, Fig.2(b) shows the almost independent
behavior of the EDM on $\varphi _{\mu }$ phase for large tan$\beta
$. In Fig.2(c) we see a slight dependence on tan$\beta $ , as we
expected for the c-quark contrary to the case of the b-quark\cite
{demvol}.

We conclude that for a reasonable portion of the parameter space
the c-quark EDM gets a gluino contribution to the order of
$10^{-21}$ e.cm.

Fig.3(a-c) show the dependence of the chargino contribution on the
parameters $\varphi _{A}$, $\varphi _{\mu }$ and tan$\beta ,$ respectively.
All the plots give a very small contribution, smaller than 10$^{-24}$ e. cm
which is at least two orders of magnitude smaller than the gluino
contribution. Although its numerical value is far below of the gluino
contribution, it would be worth mentioning the rather slight increase of the
chargino contribution with $\varphi _{\mu }$ and tan$\beta .$ In these plots
we used exact expression for ${\cal {D}}_{c}^{\chi ^{\pm }}$ which depends
on the phase of $\mu $ parameter and tan$\beta $ in a very complicated way.
In order to explain the origin of these behaviours, one needs an analytical
dependence of $\sum_{k=1}^{2}\sum_{j=1}^{2}\Im \left[ \Gamma _{\chi ^{\pm
}}^{kj}\right] $ factor on $\varphi _{\mu }$ and tan$\beta $ which could be
achieved only by using some approximations.

\bigskip

{\bf 4. CHARM QUARK EDM and }$^{1}${\bf P}$_{1}${\bf CHARMONIUM}

\bigskip

Recently the E760 collaboration\cite{armstrong} at Fermilab have announced
the first possible observation of the $^{1}$P$_{1}$ CP-odd charmonium state
in $\overline{p}p$ annihilations and the mass value of \ h$_{c}(1P)$ is
given (3525,20$\pm 0.15\pm 0.20)$ MeV. However the production of the $^{1}$P$%
_{1}$ charmonium state in the $e^{+}e^{-}$ annihilation is very interesting
from the point of view of the experimental evidence of the \ c-quark EDM,
since the coupling of photon to the h$_{c}$(1$^{1}$P$_{1}$ (1$^{+-}$))
charmonium is identical to the effective Lagrangian (1). The quantum numbers
of this CP-odd resonance coincide with those of the current density\cite
{nivkov}

\begin{equation}
J_{\alpha }(\bar{c}c|^{1}P_{1})=\bar{c}(x)\
%TCIMACRO{\overleftrightarrow{\partial }}%
_{\alpha }\gamma _{5}\ c(x)
\end{equation}

In the $e^{+}e^{-}$ annihilation the $^{1}P_{1}$state can be produced via
the $\gamma Z$ and $ZZ$ box diagrams in the framework of the SM. After using
the same consideration in Ref.\cite{demvol} one can reach the following
effective CP invariant Hamiltonian

\begin{equation}
H_{SM}=\frac{\alpha }{3\pi \sqrt{2}}G_{F}m_{e}m_{c}{\frak B}J_{\alpha }(\bar{%
c}c|^{1}P_{1})\ .\ J^{\alpha }(e^{+}e^{-}|^{1}P_{1})
\end{equation}
where the current is defined in Eq.(6), and the box function ${\frak B}$
comes from standard loop integrals \cite{veltman} and it behaves as

\begin{equation}
{\frak B\sim }\frac{1}{M_{Z}^{2}m_{c}^{2}}\ln (\frac{m_{c}}{m_{e}})
\end{equation}

In minimal SUSY with two Higgs doublets, in addition to $\gamma Z$ box
diagram there is another CP-odd Higgs scalar, A$^{0}$ which also contributes
to the formation of $^{1}P_{1}$ resonance in $e^{+}e^{-}$ annihilation.
Replacing the Z boson by $A^{0}$ in Fig 4, the SUSY contributions to the
CP-conserving effective Hamiltonian can be written

\begin{equation}
H_{SUSY}=H_{SM}[M_{Z}\longleftrightarrow M_{A^{0}}]
\end{equation}

which is typically smaller than the SM amplitude for $M_{A^{0}}> M_{Z}$.
In this sense size of the CP--conserving transitions is fixed by the
SM not by the SUSY contribution. This is an important difference
between the charm and bottom $^{1}P_{1}$ production in $e^+e^-$ collisions.

In addition to the CP-conserving decay modes, the $^{1}P_{1}$
state can also be produced via the c-EDM in $e^{+}e^{-}$ channel
which is a CP-violating mode. Grey blob in Fig.4(b) represents the
effective Lagrangian (1). In this
CP-violating mode, $e^{+}e^{-}$ system is in the most probable state $%
^{3}S_{1}$. Therefore the effective Hamiltonian violating CP\ from
the Fig. 4(b) is

\begin{equation}
H_{SUSY}=\frac{4\pi \alpha }{M_{h_{c}}^{2}}(\frac{{\cal {D}}_{c}}{e}%
)J_{\alpha }(\bar{c}c|^{1}P_{1})\ .\ (e^{+}(x)\ \gamma _{\alpha }\ e^{-}(x))
\end{equation}
which is a pure SUSY effect. The comparison between the sizes of this
CP--violating transition with the CP--conserving one is a comparison
between the SUSY and SM contributions. In fact, the CP--violating
production amplitude dominates the CP--conserving one provided that the
EDM of charm quark exceeds the critical value
\begin{equation}
\left| \frac{{\cal {D}}_{c}^{crit}}{e}\right| \sim \frac{G_{F}m_{e}}{12\sqrt{%
2}\pi ^{2}}\frac{M_{h_{c}}^{2}}{M_{Z}^{2}}\ln \frac{m_{c}}{m_{e}}\sim
10^{-26}\ {\rm cm},
\end{equation}
which is one order of magnitude smaller than the one for the $b$ quark EDM.
If ${\cal {D}}_{c}$ is large enough compared to ${\cal {D}}_{c}^{crit}$ then
the CP--violating transition Eq. (9) can be suited to generate the $h_{c}$
meson. In the preceding section we performed a scanning of the SUSY
parameter space within the bounds mentioned before, and found that ${\cal {D}%
}_{c}$ is well above the critical value Eq.(10).

Now we discuss our estimates with possible $h_c$ signatures at charm factories.
The bottomonia and charmonia production in lepton and hadron colliders
have been the primary step for experimental investigation of $B$ and $D$ mesons.
Indeed, several experiments like the $B$ factories (BABAR, CLEO and KEK-B $e^-e^+$ colliders running at
$\Upsilon(4S)$, the $D$ factories (CLEO-C and BES $e^-e^+$ colliders running
at $J/\psi$, $\psi^{\prime}$, $\psi^{\prime\prime}$ and $\psi(4140)$ resonances) as well as
$p\overline{p}$ (FNAL E789 at $\sqrt{s}=800\ {\rm GeV}$,
and CERN WA102 at $\sqrt{s}=450\ {\rm GeV}$) and $p Si$
(FNAL E771 at $\sqrt{s}=800\ {\rm GeV}$) colliders
form an arena where different bottomonia and charmonia can be produced in intermediate
and final states.

Among all $b\overline{b}$ and $c\overline{c}$ resonances we are
particularly interested in the singlet P--wave states, $h_b(nP)$
and $h_c(nP)$, as these are CP--odd states their production and
decays are highly sensitive to sources of CP violation in the
underlying theory. Presently, there is no experimental evidence
for such resonances except for the discovery of $h_c(1P)$ at FNAL
E760 experiment \cite{armstrong}. However, as the recent work
\cite{demvol} suggests, the production of the singlet P--wave
mesons (of $b\overline{b}$ type) at $e^+ e^-$ colliders are
particularly interesting in that there are spectacular
enhancements in CP--violating transitions if the underlying theory
admits large enough electric dipole moments (EDM) for heavy
quarks. In this sense, production of such resonances is
particularly sensitive to the CP violation sources beyond the
standard CKM picture as the single fermion EDMs in the standard
model (SM) are highly suppressed (See \cite{shahida} and
references therein). From the experimental point of view, $e^+
e^-$ colliders running at $\sqrt{s}\sim 2 m_b$ ($\sqrt{s}\sim 2
m_c$) form the basic environment where $h_b(nP)$ ($h_c(nP)$)
resonances can be directly produced. Therefore, the $B$ ($e.g.$
BABAR) and $D$ ($e.g.$ CLEO-C \cite{cleo-c} and \cite{bes})
factories provide the requisite experimental opportunities. Unlike
the $B$ factories where the $e^+ e^-$ center--of--mass energy is
fixed to $\Upsilon(4S)$ resonance, in $D$ factories, especially
CLEO-C \cite{cleo-c}, has a variable center--of--mass energy
enabling one to wander energies around $\sqrt{s}\sim 2 m_c$ where
$h_c(1P)$ can be directly produced.

Considering CLEO-C for definiteness and its variable center--of--mass energy, it is
possible to estimate to what extent it can observe $h_c$ meson. Let us suppose that
CLEO-C observed resonance at or in close vicinity of the E760 value of $M_{h_c}$ then
the cross section for this event will be
\begin{eqnarray}
\sigma(e^+ e^- \rightarrow h_c)=27 \left|\frac{{\cal{D}}_c}{e}\right|^{2} \left|\frac{R_{P}^{'}(0)}{R_S(0)}\right|^{2} \sigma(e^+e^-\rightarrow
^{3}S_{1}) \sim 27 \left|\frac{M_{h_c} {\cal{D}}_c}{e}\right|^{2} \sigma(e^+e^-\rightarrow ^{3}S_{1})
\end{eqnarray}
where $R_{P}$ and $R_S$ are the wavefunctions of $h_c$ and
$^{3}S_{1}$ states, respectively. Clearly, when ${\cal{D}}_c/e\sim
M_{h_c}^{-1}$ the two cross sections will be of similar size
making $h_c$ highly observable. However, except for FNAL E760, no
experiment has been able to detect such a CP--odd meson so far.
Therefore, its production cross section must be rather small
compared to $\sigma(e^+e^-\rightarrow ^{3}S_{1})$. This rareness
of $h_c$ production is directly tied up to the smallness of its
production cross section. Indeed, for the face value of
${\cal{D}}_c/e\sim 10^{-21} {\rm cm}$ as follows from the dominant
gluino production, one expects $\sigma(e^+ e^- \rightarrow
h_c)\sim 10^{-12} \sigma(e^+e^-\rightarrow ^{3}S_{1})$. If CLEO-c
observes this resonance with a much larger rate then one will need
other sources of CP violation to enhance the charm quark EDM. On
the other hand, if the observed cross section turns out to be this
size then it will clearly be an indication for SUSY CP violation.
\bigskip

{\bf 5.LIGHT GLUINO CASE}

\bigskip

It is useful to look for regions of the SUSY parameter space where
$h_c$ production cross section in other words the EDM of charm
quark is substantially enhanced. This would be useful, at least,
for excluding certain portions of the SUSY parameter space
depending on the size of the signal in near-future experiments
like CLEO-c. It is straightforward to observe that the charm quark
EDM, especially the gluino contribution, is grossly enhanced in
regions of the SUSY parameter space where the gluino and lighter
scharm are nearly degenerate and close to the hadronic scale (See
\cite{berger} for analogous discussion of the scalar bottom
quarks). To prevent any conflict with the LEP results, we must
suppress $Z$ boson couplings to scharms \cite{carena} hence we
take
\begin{eqnarray}
\sin\theta_{\widetilde{c}}=\sqrt{\frac{4}{3}} \sin\theta_{W},
\end{eqnarray}
and $M_{\widetilde{c}_1}\gg M_{\widetilde{c}_2}$. Here $\theta_{\widetilde{c}}$ is the mixing angle of scalar
charm quarks, and its is fixed by requiring that $Z$ boson does not couple to light
scharm, $\widetilde{c}_2$. On the other hand, the two scalar charms must be well splitted in
mass to suppress $e^{+}e^{-}\rightarrow Z^{\star}\rightarrow \widetilde{c}_2 \widetilde{c}_2$ and $\widetilde{c}_2
\widetilde{c}_1$ signals at LEP. Under these conditions, the gluino contribution to the charm
quark EDM becomes,
\begin{eqnarray}
\left( \frac{{\cal {D}}_{c}}{e}\right) ^{\tilde{g}}\approx \frac{\alpha_s(m_{\widetilde{g}})}{27 \pi}\
\frac{\sin\varphi_{A_c}}{m_{\widetilde{g}}}
\end{eqnarray}
when $m_{\widetilde{g}}\sim M_{\widetilde{c}_2}$. Therefore, when
$m_{\widetilde{g}}$ value lies between $m_{b}$ and $2m_{b}$ the
charm quark EDM is around $10^{-17}\ cm. \sin\varphi_{A_c}$ which
is nine orders of magnitude larger than the critical value
computed in Sec. 4. It is clear that, with such an enhancement the
$h_c$ production cross section obeys roughly $\sigma(e^+ e^-
\rightarrow h_c)\approx 10^{-4}\ \sigma(e^+e^-\rightarrow
^{3}S_{1})$ which must be in the experimentally detectable range.
With a heavy SUSY spectrum, as investigated in previous sections,
the $h_c$ production cross section is rather small, and unless the
detector has a good energy resolution it is difficult to observe
an $h_c$ resonance in $s$ channel. However, a SUSY spectrum with
light gluinos and scalar charm quark (with no conflict with the
existing experiments) it is possible to enhance the $h_c$
production cross section significantly making it easier to search
for such resonances at, for instance, the CLEO experiment.

For making our results more transparent to experimental investigation, it is useful to discuss the
quantity
\begin{equation*}
R=L\sigma (e^{+}e^{-}\rightarrow h_{c})\sim 27\times
N_{J/\psi} \left|\frac{M_{h_c} {\cal{D}}_c}{e}\right|^{2}
\end{equation*}
where $N_{J/\psi}\approx 10^{9}$ is the expected number of events at CLEO-c. For a heavy
sparticle spectrum, $R\sim 10^{-2}$ whereas for a scenario with light gluinos and
scharms one has $R\sim 10^{5}$ which is well inside the observable range provided
that $10^{9}$ $J/\psi$ mesons are produced.

\bigskip

{\bf 6. DISCUSSIONS AND CONCLUSION}

\bigskip

We have discussed the production of singlet $P$--wave charmonia
at $e^+e^-$ colliders. This process is determined
mainly by the CP--violating transitions shown in Fig. 4 (b). It is
clear that the EDM of the charm quark is the basic machinery to
generate $h_{c}$, and it is only via SUSY with explicit CP
violation that the CP--violating transition dominates the
CP--conserving channel.

The channel we discuss here, which can be directly tested at the CESR--c
collider, suggests a direct access to the SUSY phases. The amount of CP
violation provided by SUSY is large, and hard to realize in other extensions
$e.g$ two--doublet models. Therefore, any evidence for $h_c$ at the CESR--c
experiment will form a direct evidence for SUSY in general, and SUSY CP
violation in particular.

\bigskip {\bf Acknowledgements}\newline
The authors would like to thank the Scientific and Technical Research
Council of Turkey (TUBITAK) for the support. \bigskip

\newpage{}

\newpage {}
\begin{figure}[tbp]
\centering \epsfig{file=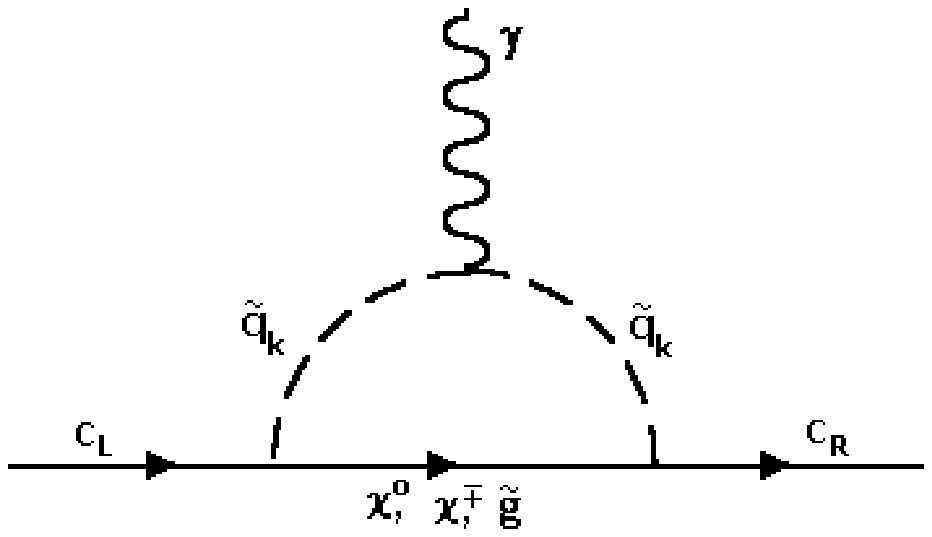,width=8cm,height=6cm}\newline
(a)
\par
\epsfig{file=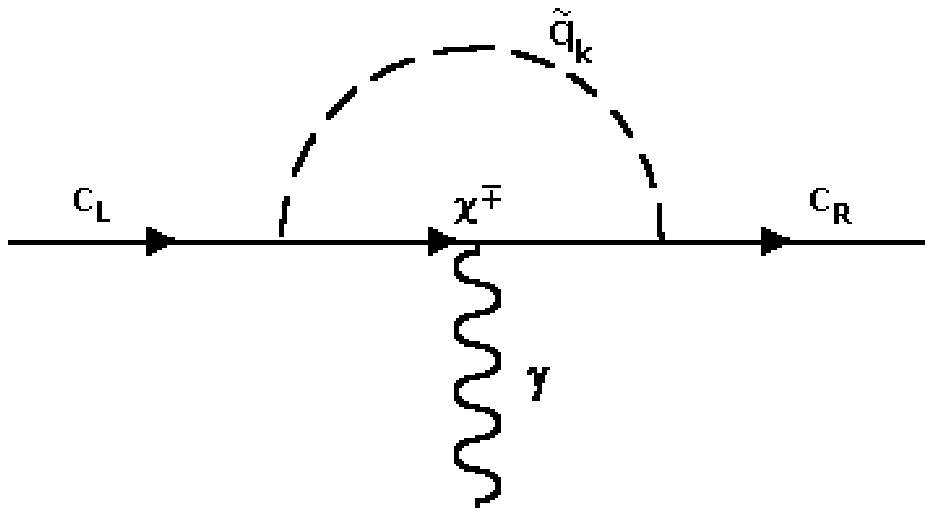,width=8cm,height=6cm}\newline (b)
\caption{The Feynman one-loop diagrams of the SUSY contributions
to the EDM of c-quark.}
\end{figure}

\newpage
\begin{figure}[tbp]
\centering\epsfig{file=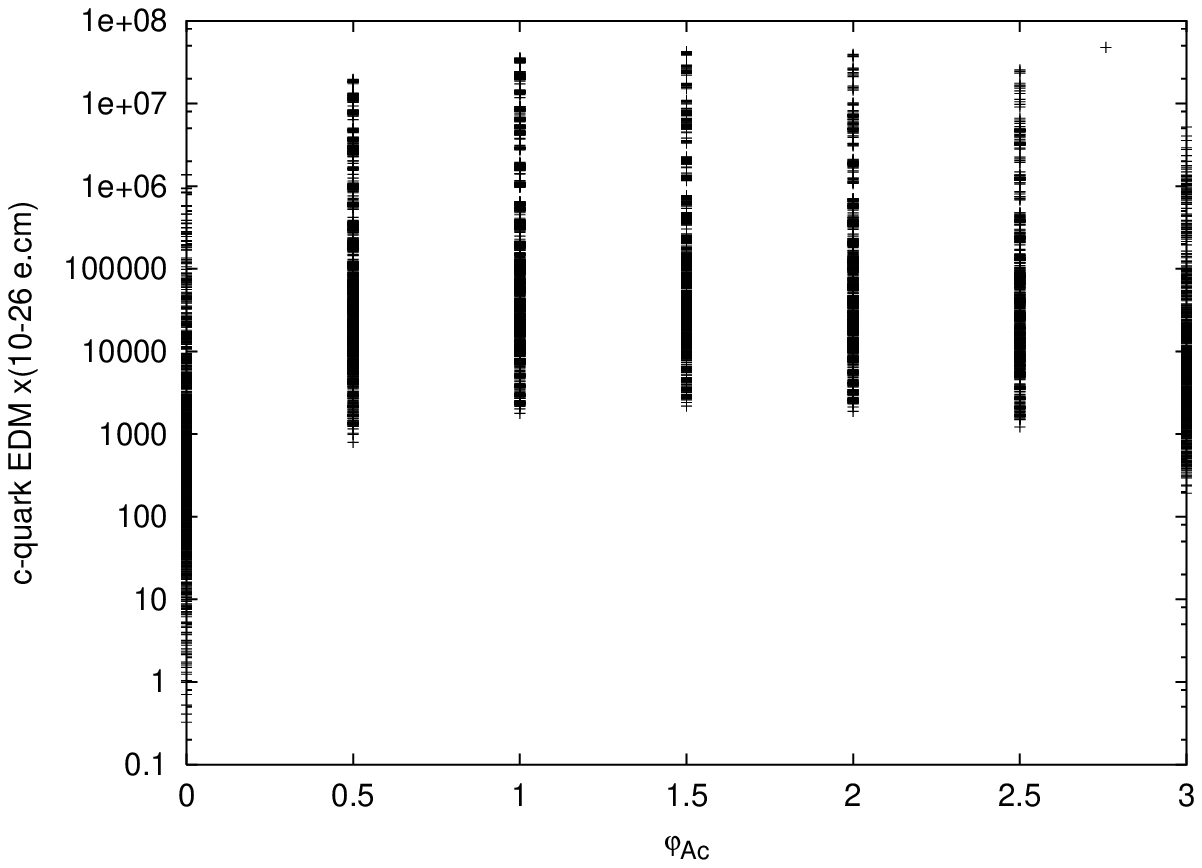,width=8cm,height=6cm}\newline (a)
\par
\centering\epsfig{file=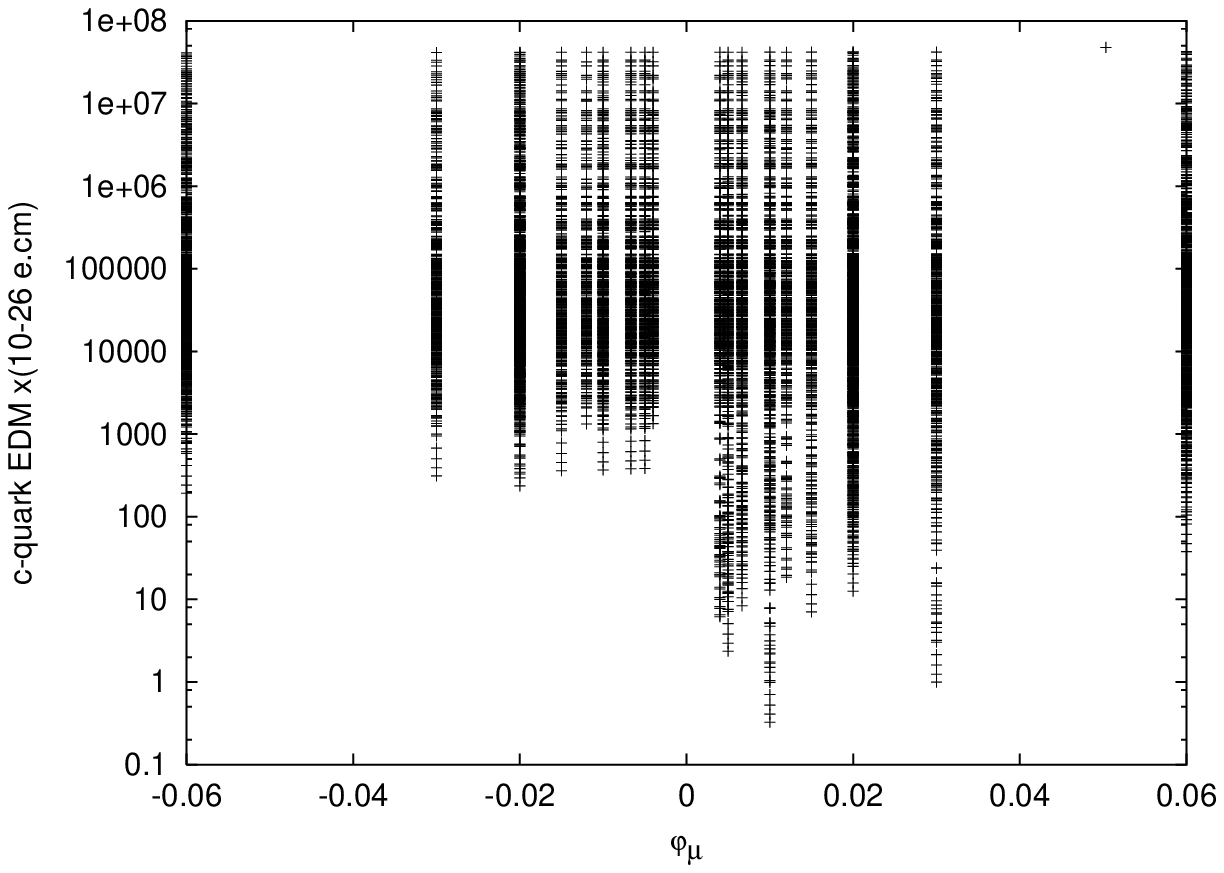,width=8cm,height=6cm}\newline (b)
\par
\centering\epsfig{file=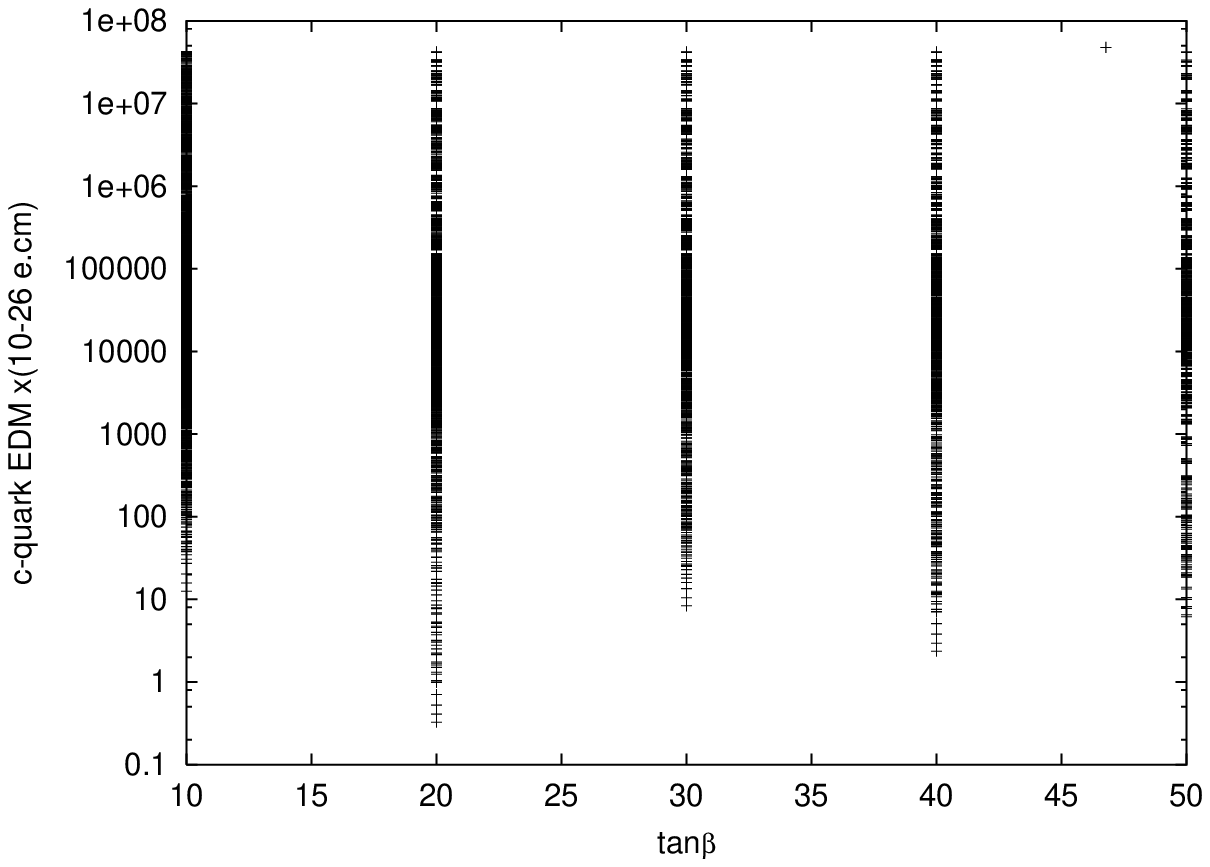,width=8cm,height=6cm}\newline (c)
\caption{The gluino contribution to c-quark EDM in units of
$10^{-26}e.cm$
as a function of {\bf {(a)}} trilinear couplings $A_{c}$ phase, {\bf {(b)}} $%
\protect\mu $ parameter phase and {\bf {(c)}} tan$\protect\beta$.}
\end{figure}

\begin{figure}[tbp]
\centering\epsfig{file=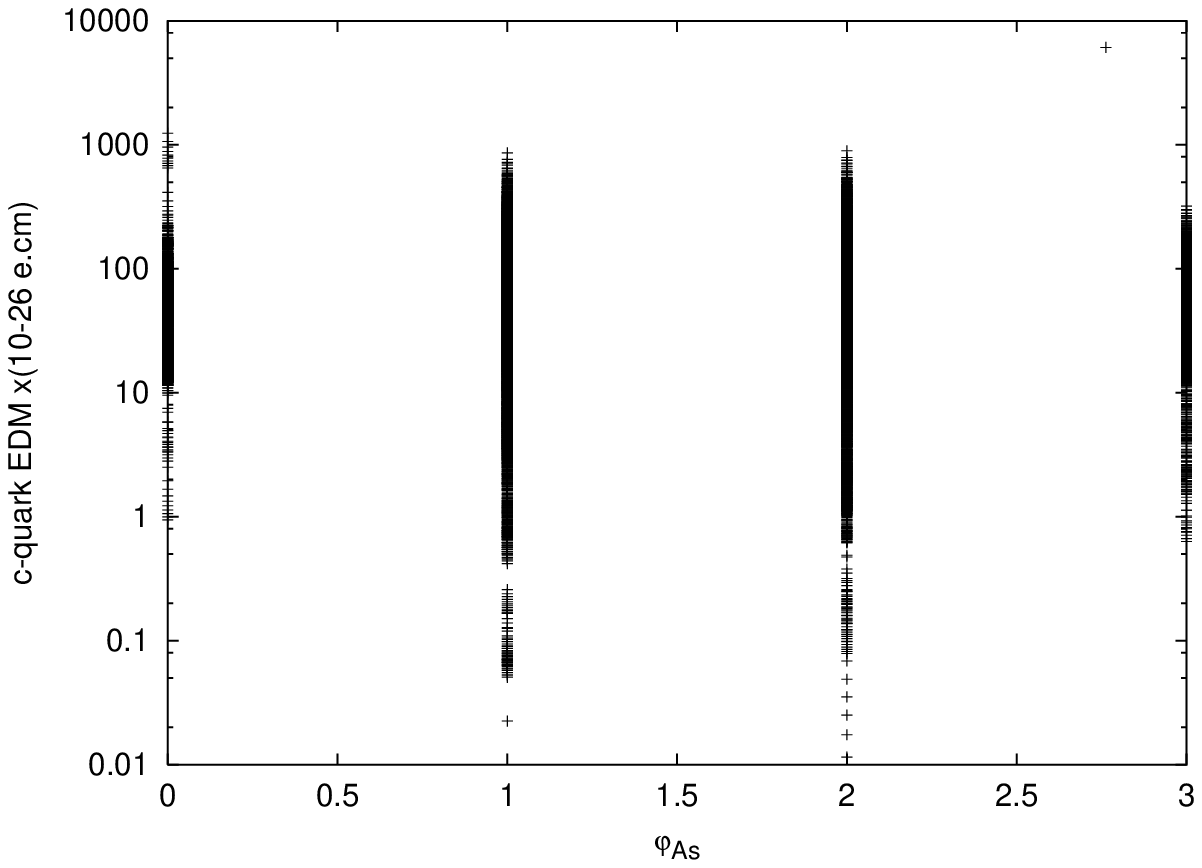,width=8cm,height=6cm}\newline (a)
\par
\centering\epsfig{file=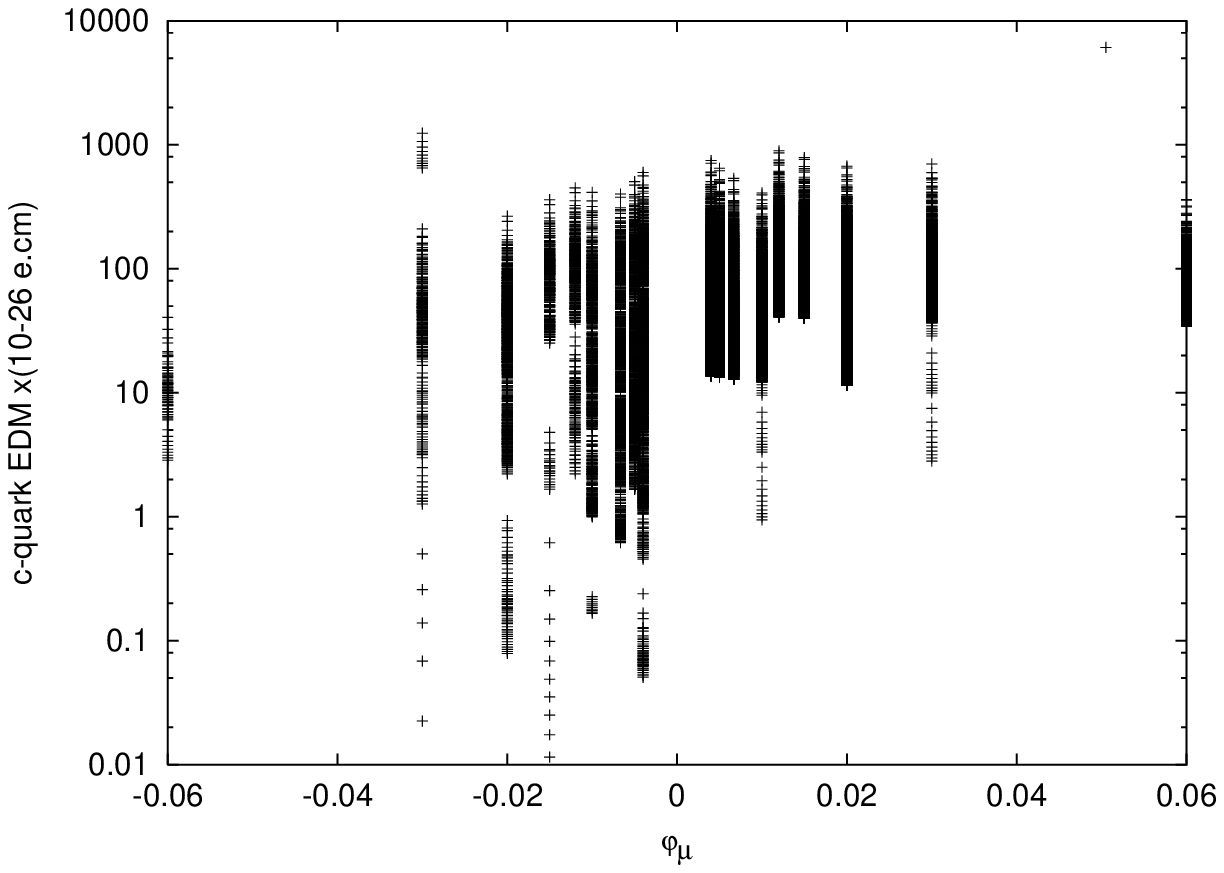,width=8cm,height=6cm}\newline (b)
\par
\centering\epsfig{file=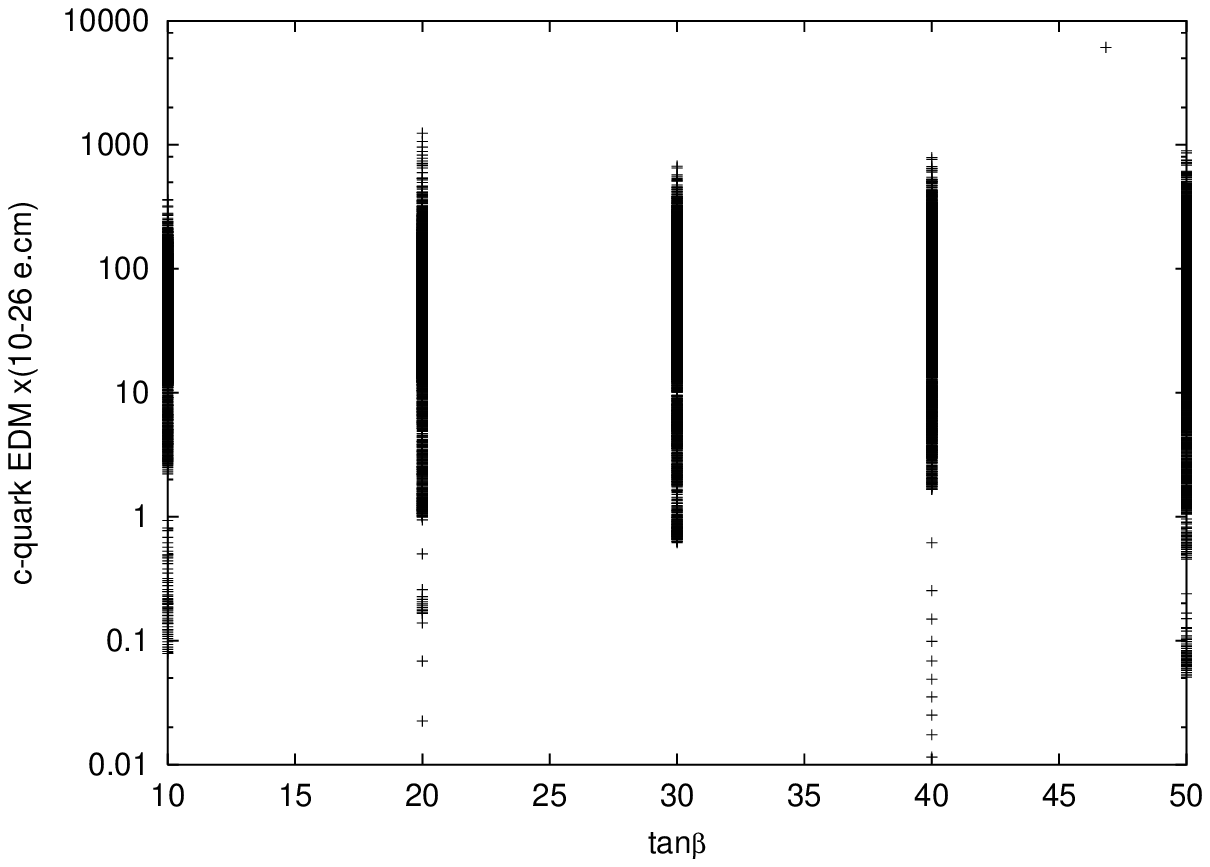,width=8cm,height=6cm}\newline (c)
\caption{ The chargino contribution to c-quark EDM in units of
$10^{-26}e.cm$
as a function of {\bf {(a)}} trilinear couplings $A_{s}$ phase, {\bf {(b)}} $%
\protect\mu $ parameter phase and {\bf {(c)}} tan$\protect\beta$.}
\end{figure}
\newpage
\begin{figure}[tbp]
\centering\epsfig{file=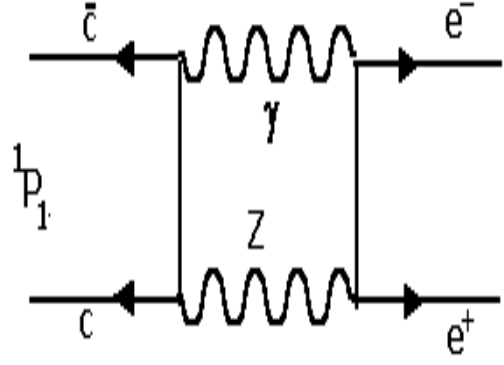,width=8cm,height=6cm}\newline (a)
\par
\centering\epsfig{file=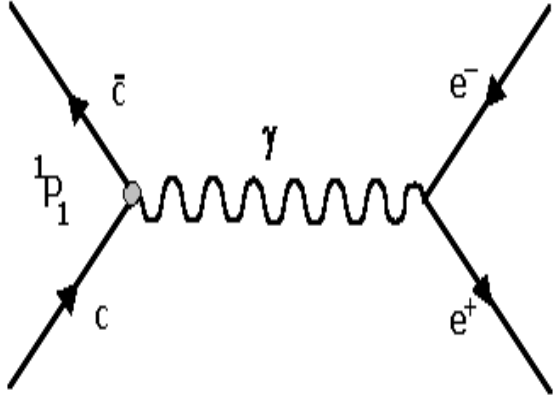,width=8cm,height=6cm}\newline (b)
\caption{The Feynman diagrams of the $^{1}$P$_{1}$ charmonium resonance in $%
e^{+}e^{-}$ scattering: {\bf {(a)}} The CP-conserving decay mode and {\bf {%
(b)} }is the CP-violating decay mode where the grey blob stands for the
insertion of the effective Lagrangian (1).}
\end{figure}

\end{document}